# Truthful Equilibria in Generalized Common Agency Models

Ilias Boultzis*

**Abstract**

In this paper I discuss truthful equilibria in common agency models. Specifically, I provide general conditions under which truthful equilibria are plausible, easy to calculate and efficient. These conditions generalize similar results in the literature and allow the use of truthful equilibria in novel economic applications. Moreover, I provide two such applications. The first application is a market game in which multiple sellers sell a uniform good to a single buyer. The second application is a lobbying model in which there are externalities in contributions between lobbies. This last example indicates that externalities between principals do not necessarily prevent efficient equilibria. In this regard, this paper provides a set of conditions, under which, truthful equilibria in common agency models with externalities are efficient.

## 1 Introduction

The common agency model is a game in which many principals share a common agent. Economists apply this model in many areas of economic research like lobbying, industrial organization, public economics e.t.c.

Common agency models often have many equilibria. However, truthful equilibria are probably the most popular among them. This type of equilibria was introduced by Bernheim and Whinston (1986) for quasi linear utility functions and was generalized to all utility functions by Dixit et al. (1997). Specifically, Dixit et al. (1997) argued that truthful equilibria are focal because they have three important properties: they are easy to calculate, Pareto efficient and plausible. The last property means that the best response set of the principals always contains a strategy which is consistent with truthful equilibria (truthful strategy).

*Department of Economics, Athens University of Economics and Business, e-mail: boultzis@gmail.com. I would like to thank the participants of the 2019 CRETE conference for their valuable comments. I would also like to thank Apostolis Philippopoulos for our discussions, from an early stage of this paper and Katerina Katsandredaki for her assistance in the editing of this paper. The usual disclaimer applies.

However, the analysis by Dixit et al. (1997) was motivated by lobbying games. Following this motivation, they based their results on a general setting associated with lobbying models. This setting consists of three key assumptions. First, the utility of the principals (lobbyists) decreases with the bids (contributions) they offer to the agent (politician). Second, the utility of the principals depends only on their own bids. Third, the utility of the agent is increasing in all bids offered by the principals.

Nevertheless, many applications of common agency do not fit this restricted framework. Thus, a question arises. Is it possible to find a broader set of conditions under which truthful equilibria can be used? In this paper, I attempt to answer this question. Specifically, I provide general conditions under which the results of Dixit et al. (1997) survive. These conditions allow the use of truthful equilibria in novel economic applications.

In this paper I consider two such applications. The first application is a market game in which a group of principals (sellers), sells a uniform good to a single agent (buyer). The sellers move first and present the buyer with a bid (price) which is conditioned on the amount that the buyer wishes to buy. Then, the buyer decides on the quantity he buys from each seller. This situation is the reverse of the lobbying model by Dixit et al. (1997), since typically, the utility of the sellers increases, while the utility of the buyers decreases following a rise in prices. The properties of truthful equilibria in such market games were established by Bernheim and Whinston (1986) for quasi linear utility functions. Here I follow Dixit et al. (1997) and generalize them to all utility functions.

The second application considers externalities in bids, among principals. These externalities occur naturally when principals are interrelated in other ways besides sharing a common agent. For example, think of a federal country in which the states (principals) lobby the central government (agent) for a transfer. Moreover, assume that the states finance voluntarily a public good like public safety. In this case, all states benefit from public safety spending in the other states. However, this spending depends on the available resources of each state. In turn, these available resources decrease with lobbying. Thus, the welfare in each state depends on lobbying expenditures in all states. Therefore, externalities in bids, in this case spending for lobbying, emerge.

The use of truthful equilibria in the two applications above is not equally intuitive. Specifically, in market games the setting resembles the model of Dixit et al. (1997). Thus, the fact that truthful equilibria retain their properties does not come as a surprise. However, externalities in bids is a different story. In such models the literature has identified efficiency failures[1] and thus the efficiency of truthful equilibria is less expected. In particular, externalities in bids can lead to inefficiency because of a possible prisoners' dilemma. The set of conditions I provide in this paper imply that this prisoners' dilemma need not always appear. Thus, there exist applications with externalities in bids, in which truthful equilibria have all three properties identified by Dixit et al. (1997), including efficiency.

In this regard, my model belongs in the strand of literature that explores the robustness of truthful equilibria to various extensions of the original common agency model. Dixit et al. (1997) pioneered this literature by extending the model of Bernheim and Whinston (1986) to general utility functions. Other important examples of this literature include Bergemann and Välimäki (2003) who consider dynamic common agency, Prat and Rustichini (2003) who discuss multiple agents and Martimort and Stole (2009b) who consider asymmetric information. Furthermore, my paper relates to the literature discussing common agency with externalities in bids. Examples of this literature are Peters (2001), Martimort and Stole (2002), Peters and Szentes (2012), Szentes (2014) and Galperti (2015). The papers in this strand of literature discuss models with asymmetric information and observe that externalities in bids can lead to inefficient equilibria. Following this observation, they allow bids to depend on the bidding strategies of the other principals in order to restore efficiency. These papers relate to my work, to the extent that their results also apply to symmetric information models. In this respect, my contribution is that truthful equilibria can be efficient, even if bids do not depend on other bidding strategies.

---

[1] See Peters (2001) and Martimort and Stole (2002).

The rest of the paper is organised as follows. Section 2 describes the main model, section 3 presents the key results and section 4 investigates a variation of the main model. Section 5 discusses the general issue of efficiency of equilibria and the relationship between my paper and the existing literature on externalities. Section 6 considers economic applications and section 7 concludes. Finally, appendix A contains the main proof of the paper, while appendix B, which is not intended for publication, contains the rest of the proofs along with some examples and calculations.

## 2 Model

### 2.1 Setting

Before I continue with the details of the model, let me introduce the notation that I use in the rest of the paper. The index $i$ runs from 1 to $n$. Furthermore, except when otherwise stated, I use Latin and Greek letters in the following manner. Consider for example the lower-case letter "$x$". Then, $x_i$ is a real number, $x$ is the vector of all $x_i$, $x_{-i}$ is the vector containing all members of $x$ except $x_i$, $\tilde{x} = \sum_i x_i$ and $\tilde{x}_{-j} = \sum_{i \neq j} x_i$. Moreover, I use the symbol $x_i(\cdot)$ to describe a function $x_i : Z \to R$, such that $x_i = x_i(z)$, for a given set $Z$. The symbols $x(\cdot)$ and $x_{-i}(\cdot)$ describe the respective vectors of functions. Finally, if $x$, $y$ are two vectors, then $x \geq y$ means that $x_i \geq y_i$ for all $i$, while $x > y$ means that $x_i \geq y_i$ for all $i$ and there exists at least one $i$, such that $x_i > y_i$.

I turn now to the model. Consider a common agency model with one agent and $n$ principals. Following Dixit et al. (1997) I discuss here what is known as a public common agency model[2]. I depart from this assumption in section 4.

---

[2] In public common agency models the principals condition their bids on the entire action of the agent, while in private common agency, the principals

***Agent.*** The agent chooses an element $a$ from the set $A$. The set $A$ reflects budget, institutional or other constraints that depend on specific applications. For example, if the agent is a government choosing a tax rate, the set $A$ is the interval $[0, 1]$. Likewise, if the agent is a buyer who is buying a quantity of a good from each of the principals, the set $A$ is a subset of $R_+^n$. Henceforth, I refer to $a$ as the agent's action.

The utility function of the agent is a function:

$$u_0 : A \times R^n \to R \quad \text{such that} \quad u_o = u_0(a, b).$$

The vector $b \in R^n$, is the vector of bids that the principals submit to the agent in order to influence the choice of $a$. This utility function is strictly monotonous with respect to all bids and continuous with respect to all its elements.

***Principals.*** On the other hand, each principal chooses a bidding function:

$$b_i : A \to R \quad \text{such that} \quad b_i = b_i(a),$$

in order to influence the agent. These bidding functions meet appropriate restrictions. Specifically, there exist two functions $\overline{b_i} : A \to R$ and $\underline{b_i} : A \to R$, which are uniformly bounded above and below by $b_{max}, b_{min} \in R$ respectively and satisfy the inequality $\overline{b_i}(a) \geq \underline{b_i}(a)$ for all $a \in A$. These functions define feasible bids:

**Definition 1.** ***Feasibility***: A bid $b_i \in R$ is feasible relative to $a \in A$, if $b_i \in [\underline{b_i}(a), \overline{b_i}(a)]$. Moreover, a bidding function $b_i(\cdot)$ is feasible, if $b_i(a)$ is feasible relative to $a$, for all $a \in A$.

Additionally, the vector $b \in R^n$ is feasible relative to $a \in A$ if all $b_i$ are feasible relative to $a$. In this case I say that the pair $(a, b)$ is feasible. A feasible pair $(a, b)$ is **symmetric** if $b_i = b_j$ for all $i, j$. Similarly, the vector of bidding functions $b(\cdot)$ is feasible, if all its elements are feasible. Moreover, if $b(\cdot)$ is feasible and $a \in A$, I say that $(a, b(\cdot))$ and $(a, b_i(\cdot))$ are feasible. A feasible pair $(a, b(\cdot))$ is **symmetric**

---
condition their bids only on a part of the agent's action (i.e. the part they observe). For more on the meaning of these terms see Martimort and Stole (2009a).

if $b_i(\cdot) = b_j(\cdot)$ for all $i,j$.

Feasibility restrictions in bids reflect application specific constraints. For example, if the principals are lobbies offering campaign contributions, the bids must be positive and not exceed the budget constraint of the lobby. Likewise, if the principals are sellers and the bids are selling prices, the bids should be greater than the cost of acquisition and smaller than the buyer's reservation price.

Now I turn to the utility functions of the principals. Specifically, the utility function of principal $i$ is a function:

$$u_i : A \times R^n \to R \quad \text{such that} \quad u_i = u_i(a,b).$$

This utility function is strictly monotonous with respect to own bids $b_i$ and continuous with respect to both $a$ and $b$.

**Timing.** Finally, the timing of the model is standard. There are two stages. In stage one, the principals submit simultaneously their bidding functions. In stage two, the agent chooses his action and the bids are realised.

Following the analysis above, my model is fully described by the number of principals $n$, the $n+1$ utility functions, the set $A$ and the restrictions in bids $(\underline{b_i}(\cdot), \overline{b_i}(\cdot), b_{max}, b_{min})$. Henceforth, I use the term game, whenever I refer to a common agency model defined in such a way.

This game extends the model by Dixit et al. (1997) in two ways. First, the utility of each principal also depends on the bids of all other principals. Thus, my model allows for externalities among principals. Second, I make no prior assumptions, regarding the effect of bids on utility, other than monotonicity. These extensions broaden the range of applications to which truthful equilibria apply. I discuss these issues again in 2.4 which provides further specification of the model and in section 6 which discusses economic applications.

Let me now turn to the equilibrium of the game.

## 2.2 Equilibrium

The definitions of best response and equilibrium below extend the respective definitions by Dixit et al. (1997) [3].

**Definition 2.** *Best response*: A feasible bidding function $b_i(\cdot)$, belongs in the best response set of principal $i$, to the feasible bidding functions $b_{-i}(\cdot)$ of the other principals, if:
There exists an $a' \in arg \max_{a \in A} u_0(a, b(a))$, such that there does not exist a feasible pair $(a^*, b_i^*(\cdot))$, such that $u_i(a^*, b_i^*(a^*), b_{-i}(a^*)) > u_i(a', b(a'))$ and $a^* \in arg \max_{a \in A} u_0(a, b_i^*(a), b_{-i}(a))$.

**Definition 3.** *Equilibrium*: A feasible pair $(a^o, b^o(\cdot))$ is an equilibrium if:
**a)** $a^o \in arg \max_{a \in A} u_0(a, b^o(a))$ and
**b)** for all $i$, there does not exist a feasible pair $(a', b_i(\cdot))$, such that $a' \in arg \max_{a \in A} u_0(a, b_i(a), b_{-i}^o(a))$ and $u_i(a', b_i(a'), b_{-i}^o(a')) > u_i(a^o, b(a^o))$.

Let me now turn to the notion of truthful equilibrium.

## 2.3 Truthful equilibrium

Truthful equilibria refine the equilibria described in definition 3. They were introduced by Bernheim and Whinston (1986) and were generalized by Dixit et al. (1997). In truthful equilibria the principals submit truthful bidding functions. These functions exactly reflect changes in the utility of the principals that follow from changes in the actions of the agent. Thus, truthful bidding functions reveal the true preferences of the principals. Definitions 4 and 5 below adapt these ideas to my setting.

Let $(a, b(\cdot))$ be a feasible pair and $u_i^* \in R$. Consider the equation $u_i^* = u_i(a, \phi_i, b_{-i}(a))$, with respect to $\phi_i$. Since $b(\cdot)$ is a vector of feasible bidding functions, $b_{-i}(a)$ exists for all $a \in A$. Furthermore, because the utility of the principals

---
[3] On the definitions 2 and 3 see also Ko (2011) and Ko (2017).

is monotonous in own bids, this equation always has a unique solution. This solution defines a function $\phi_i : A \to R$ such that $\phi_i = \phi_i(a; u_i^*, b_{-i}(\cdot))$. Then, I define truthful responses as follows:

**Definition 4.** *Truthful response*: A bidding function $b_i^T : A \to R$, is a truthful response of principal $i$ to the feasible bidding functions $b_{-i}(\cdot)$ of the other principals, relative to the constant $u_i^*$, if :

a)
$$b_i^T = \begin{cases} \underline{b_i}(a) & \text{if } u_i(a, \underline{b_i}(a), b_{-i}(a)) < u_i^* \\ \phi_i(a; u_i^*, b_{-i}(\cdot)) & \text{if } u_i(a, \overline{b_i}(a), b_{-i}(a)) \leq u_i^* \leq u_i(a, \underline{b_i}(a), b_{-i}(a)) \\ \overline{b_i}(a) & \text{if } u_i^* < u_i(a, \overline{b_i}(a), b_{-i}(a)) \end{cases}$$

and $u_i(\cdot)$ is strictly decreasing in own bids, or

b)
$$b_i^T = \begin{cases} \underline{b_i}(a) & \text{if } u_i(a, \underline{b_i}(a), b_{-i}(a)) > u_i^* \\ \phi_i(a; u_i^*, b_{-i}(\cdot)) & \text{if } u_i(a, \overline{b_i}(a), b_{-i}(a)) \geq u_i^* \geq u_i(a, \underline{b_i}(a), b_{-i}(a)) \\ \overline{b_i}(a) & \text{if } u_i^* > u_i(a, \overline{b_i}(a), b_{-i}(a)) \end{cases}$$

and $u_i(\cdot)$ is strictly increasing in own bids.

Definition 4 states that truthful responses are equal to the expression $\phi_i(a; u_i^*, b_{-i}(\cdot))$ except when this expression violates lower or upper feasibility bounds. In such cases the truthful responses are equal to these bounds. Therefore, truthful responses are by construction feasible bidding functions[4].

Now I can turn to the definition of truthful equilibrium.

**Definition 5.** *Truthful equilibrium*: Let $(a^o, b^o(\cdot))$ be an equilibrium of the game and $u^o = u(a^o, b^o(a))$ be the vector of equilibrium utility levels of the principals. This equilibrium is truthful, if for all $i$, the equilibrium bidding function $b_i^o(\cdot)$ is a truthful response of principal $i$ to the equilibrium bidding functions $b_{-i}^o(\cdot)$ of the other principals, relative to his equilibrium utility level $u_i^o$.

---

[4] Ko (2011) explains the advantages of definition 4 when compared to $b_i^T = min\{\overline{b_i}(a), max\{\underline{b_i}(a), \phi_i(a; \overline{u}_i, b_{-i}(\cdot))\}\}$ which corresponds to the definition provided by Dixit et al. (1997).

Let me clarify the notion of truthful equilibrium. Consider any feasible pair $(a, b(\cdot))$ and let $u = u(a, b(a))$, be the vector of corresponding utility levels. Then, definition 4 determines the truthful response, for each principal, to the bidding functions $b_{-i}(\cdot)$ of the other principals, relative to $u_i$. For a given $a$, this operation defines a mapping from the n-dimensional space of feasible bidding functions to itself. Now consider an equilibrium pair $(a^o, b^o(\cdot))$ and let $u^o = u(a^o, b^o(a^o))$. If $b^o(\cdot)$ is a fixed point in the mapping above, then it is a truthful equilibrium[5].

Next, I discuss the structure of the game.

## 2.4 Additional assumptions

Here, I provide some additional assumptions that often characterize applications.

**Assumption A.** *Opposing monotonicity*

**A1. Lobbying.** The utility of all principals is strictly decreasing in own bids and the utility of the agent is strictly increasing in all bids. Moreover, $\underline{b}_i(a) = b_{min}$, for all $i$ and $a \in A$.

---

[5] The existence of such a fixed point deals with the issue of infinite regress that can appear in common agency with externalities. For the problem of infinite regress in common agency see Peters (2001), Martimort and Stole (2002) and more recently Szentes (2014) and Galperti (2015). In these papers the bidding functions explicitly depend on other bidding functions. Therefore, infinite regress can appear because the bidding function of principal $i$ depends on the bidding function of principal $j$, which in turn depends on the bidding function of principal $i$ and so on. My setting does not allow for explicit dependence on other bidding functions. However, the principals form guesses about the other bidding functions which also depend on the guesses that the other principals form and so on.

**A2. Market.** The utility of all principals is strictly increasing in own bids and the utility of the agent is strictly decreasing in all bids. Moreover, $\overline{b}_i(a) = b_{max}$, for all $i$ and $a \in A$.

Assumption A1 is often satisfied in lobbying models. In these models the principals are lobbies and their bids are usually campaign contributions to politicians. In such a case, the politicians like receiving contributions while the lobbies dislike paying them. Moreover, contributions must be non-negative and therefore, the common lower bound $b_{min}$ is zero. Assumption A2 is more relevant in market games, in which the principals are sellers of a homogeneous good and their bids are selling prices. In this case, the sellers like high prices while the buyer dislikes them. Furthermore, the upper bound of the bids is the buyer's reservation price.

Henceforth, when I refer to assumptions A1 and A2, I use the terms lobbying and market monotonicity respectively.

**Assumption B.** *No externalities.*
The utility of the principals takes the form $u_i : A \times R \to R$ such that $u_i = u_i(a, b_i)$.

Assumption B describes a special case without externalities in bids. The combination of lobbying monotonicity and no externalities defines the lobbying model discussed by Dixit et al. (1997).

**Assumption C.** *Conflict of interests at* $(a, b)$.
Let $(a, b)$ be a feasible pair. If there exists a feasible pair $(a, b')$ such that $u_0(a, b') > u_0(a, b)$ then there exists an $i$ such that $u_i(a, b') < u_i(a, b)$ and if there exists a feasible pair $(a, b')$ such that $u(a, b') > u(a, b)$ then $u_0(a, b') < u_0(a, b)$.

Assumption C states that for a given agent's action, it is impossible to change bids, in a way that makes both the principals and the agent better off. Therefore, this assumption introduces conflict of interests between principals and agent, over bids. Assumption C is implied by opposing monotonicity in games without externalities. However, in games with externalities this is not always the case.

In order to explain the role of externalities in this issue, let me consider a game with lobbying monotonicity. First, assume that there are no externalities. Also, fix the agent's action at a certain level and consider a change in bids which makes the agent better off. Such a change requires that at least one bid ($b_i$) increases. Then, since $u_i = u_i(a, b_i)$, the utility of at least one principal decreases. Thus, these changes are consistent with conflict of interests. Alternatively, consider a similar game with externalities. In this case, the utility of principal $i$ is $u_i = u_i(a, b_i, b_{-i})$. Furthermore, assume that the utility of all principals is increasing in all the elements of $b_{-i}$. Also, assume that all contributions increase. Then, two conflicting effects emerge. On the one hand, the increase in own bids has a negative effect on the utility of the principals. On the other hand, the increase in the other bids has a positive effect on the principals. If the positive cross effect dominates the negative own effect, for all principals, then conflict of interests is violated. In such a case, an increase in all bids increases the utility of all principals and the agent. Assumption C appropriately restricts externalities to disallow such situations. This assumption along with assumption D that follows guarantee the efficiency of truthful equilibria in games with externalities.

**Assumption D.** *Deep pockets*

**D 1.** *Weak deep pockets at $(a^o, b^o(\cdot))$.*

Let $(a^o, b^o(\cdot))$ be a truthful equilibrium and $(a^*, b^*)$ be a feasible pair such that $u_0(a^*, b^*) \geq u_0(a^o, b^o(a^o))$ and $u(a^*, b^*) \geq u(a^o, b^o(a^o))$ with at least one strict inequality, then $u(a^o, b^o(a^o)) \geq u(a^*, b^o(a^*))$.

**D 2.** *Strong deep pockets*

There exists $\underline{u_i} \in R$ such that:

**D 2.1.** If the utility of all principals is strictly decreasing in own bids then $u_i(a, \overline{b_i}(a), b_{-i}(a)) = \underline{u_i} \leq u_i(a, b(a))$ for all $i$, $a \in A$ and feasible $b_{-i}(\cdot)$.

**D 2.2.** If the utility of all principals is strictly increasing in own bids then $u_i(a, \underline{b_i}(a), b_{-i}(a)) = \underline{u_i} \leq u_i(a, b(a))$ for all $i$, $a \in A$ and feasible $b_{-i}(\cdot)$.

The term "deep pockets" is due to Ko (2011) who uses a similar assumption. Furthermore, Dixit et al. (1997) also employ a version of strong deep pockets.

Specifically, Dixit et al. (1997) assume that there is a subsistence utility level $\underline{u_i}$ and define $\overline{b_i}(\cdot)$ implicitly through $u_i(a, \overline{b_i}(a)) = \underline{u_i}$. If a game satisfies strong deep pockets then it also satisfies weak deep pockets. However, as I show later on, many applications satisfy weak deep pockets directly. These applications include all games without externalities but also a number of games with externalities.

Definition 6 introduces some terms that I use in assumptions E and F below.

**Definition 6.** *Game structure:*

**a)** A game is **differentiable** if all utility functions are differentiable with respect to all bids.

**b)** A game is **cumulative** if the utility functions of the agent and the principals are as follows:

$u_0 : A \times R \to R$ such that $u_0 = u_0(a, \tilde{b})$ and

$u_i : A \times R^2 \to R$ such that $u_i = u_i(a, b_i, \tilde{b}_{-i})$ for all $i$.

**c)** A game exhibits **negative externalities** if it is differentiable, cumulative and the utility of all principals is strictly decreasing in the total bids of the other principals ($\tilde{b}_{-i}$).

**d)** A game exhibits **positive externalities** if it is differentiable, cumulative and and the utility of all principals is strictly increasing in the total bids of the other principals ($\tilde{b}_{-i}$).

**e)** A game is **symmetric** if $u_i(\cdot) = u_j(\cdot)$, $\underline{b_i}(\cdot) = \underline{b_j}(\cdot)$ and $\overline{b_i}(\cdot) = \overline{b_j}(\cdot)$ for all $i, j$.

**f)** A game is **quasi-concave** if it is cumulative and the utility functions of all principals are quasi-concave with respect to own and other bids.

**Assumption E.** *Small externalities*

Either the game exhibits negative externalities and $\left|\dfrac{\partial u_i}{\partial b_i}\right| > \left|\dfrac{\partial u_i}{\partial \tilde{b}_{-i}}\right|$ for all $i$ and all feasible $(a, b)$ or the game exhibits positive externalities and $\left|\dfrac{\partial u_i}{\partial b_i}\right| > (n-1)\left|\dfrac{\partial u_i}{\partial \tilde{b}_{-i}}\right|$ for all $i$ and all feasible $(a, b)$.

Assumption E is a special case of conflict of interests. Specifically, small externalities describe a situation in which the effect of own bids appropriately dominates the respective cross effects. In this respect, think of $(n-1)\left|\frac{\partial u_i}{\partial \tilde{b}_{-i}}\right|$ as the total cross effect and of $\left|\frac{\partial u_i}{\partial \tilde{b}_{-i}}\right|$ as the average cross effect. These two effects coincide when $n = 2$. As it turns out, this restriction in externalities achieves the conflict between agent and principals which is necessary to satisfy assumption C.

**Assumption F.** *Symmetric negative externalities*
The game is symmetric, quasi concave and exhibits negative externalities.

Proposition 1 that follows explains the relationship between assumptions A-F.

**Proposition 1.**
**(i)** Strong deep pockets imply weak deep pockets at all truthful equilibria.
**(ii)** The combination of opposing monotonicity and no externalities, implies conflict of interests at all feasible pairs $(a, b)$ and weak deep pockets at all truthful equilibria.
**(iii)** The combination of small externalities and lobbying monotonicity implies conflict of interests at all feasible pairs $(a, b)$ and weak deep pockets at all truthful equilibria.
**(iv)** The combination of symmetric negative externalities and lobbying monotonicity implies conflict of interests at all symmetric pairs $(a, b)$ and weak deep pockets at all symmetric truthful equilibria.
Proof: See appendix B.1.1. Also, Dixit et al. (1997) prove a part of (ii) by showing that lobbying monotonicity and no externalities imply conflict of intersts and weak deep pockets during their proof of the efficiency of truthful equilibria[6].

I turn now to the main results.

---
[6] For their proof see Dixit et al. (1996)

# 3   Results

Dixit et al. (1997), like Bernheim and Whinston (1986) before them, argue that truthful equilibria are focal, because they share three key properties. Namely, truthful equilibria are plausible, easy to calculate and efficient[7].

Dixit et al. (1997) arrive at this result under the assumptions of lobbying monotonicity and no externalities. Here I generalize their argument by providing a broader set of conditions under which it is valid. The four propositions that follow achieve this task.

**Proposition 2.** *Plausibility*

Consider a game that exhibits opposing monotonicity. Then, the best response set of principal $i$ to the bidding functions of the other principals $b_{-i}(\cdot)$ always contains a truthful response.

Proof: see appendix B.1.2.

**Proposition 3.** *Calculation*

Consider a game that exhibits opposing monotonicity.

**A.** If the feasible pair $(a^o, b^o(\cdot))$ is an equilibrium then (Ai) and (Aii) below are true:

**(Ai)** $a^o \in arg \max_{a \in A} u_0(a, b^o(a))$

**(Aii)** For all $i$, lobbying monotonicity implies that $u_0(a^o, b^o(a^o)) = \max_{a \in A} u_0(a, b_{min}, b_{-i}(a))$ and market monotonicity implies that $u_0(a^o, b^o(a^o)) = \max_{a \in A} u_0(a, b_{max}, b_{-i}(a))$

**B.** Let $b^o(\cdot)$ be a vector of feasible bidding functions, such that for all $i$, $b_i^o(\cdot)$ is a truthful response to $b_{-i}^o(\cdot)$, relative to $u_i^o \in R$. Also let $a^o \in A$ be an agent's action such that the pair $(a^o, b^o(\cdot))$ is feasible and satisfies (Ai) and (Aii) above. Then if (Bi) and (Bii) below are true $(a^o, b^o(\cdot))$ is a truthful equilibrium of the game.

---

[7] A part of the literature expresses doubts regarding the relevance of truthful equilibria. In this respect Kirchsteiger and Prat (2001) provide an experimental argument and Martimort and Stole (2009b) provide a theoretical argument against truthful equilibria.

**(Bi)** $u^o = u(a^o, b^o(a^o))$

**(Bii)** There does not exist an $a \in A$ such that the pair $(a, b^o(\cdot))$ is feasible, satisfies conditions (Ai) and (Aii) and yields $u(a, b^o(a)) > u^o$.

Proof: See appendix B.1.3.

**Lemma to proposition 3B**

If the game satisfies strong seep pockets then condition (Bii) is always satisfied.

Proof: See the proof of part (i) of proposition 1 in appendix B.1.1.

These results are true regardless of other characteristics of the game i.e. the existence of externalities. In more detail, proposition 2 states that the principals stand to loose nothing from responding truthfully to any bidding function chosen by the other principals. Also, the intuition behind proposition 3A is standard. In particular, principal $i$ submits a bid that matches the agent's outside option. Principal $i$ has no motive to improve his bid any further.

Proposition 3A holds for all equilibria. Proposition 3B states that the converse of 3A holds only for truthful equilibria and only under certain conditions. In order to explain how proposition 3B works let me give two examples: First, consider a game with two principals with utility functions $u_i = a - b_i$ and an agent with utility function $u_0 = b_1 + b_2$. Moreover, assume that $a \in [0, 1]$ and $b_i \in [0, a]$. This game satisfies strong deep pockets since $u_i(a, b_i) \geq 0$ for all feasible $(a, b_i)$, $\overline{b_i}(a) = a$ and $u_i(a, \overline{b_i}(a)) = 0$ for all $a \in A$. Moreover, in this game the bidding functions $b_i = 0$ for all $a \in [0, 1]$ are truthful responses to each other relative to $u_i^o = 1$ and satisfy conditions (Ai) and (Aii) for all $a \in [0, 1]$. Then, proposition 3B implies that only $a = 1$ for which $u_i = 1 = u_i^o$ is part of a truthful equilibrium with equilibrium utility $u_i^o = 1$.

Second, consider an example in which condition (Bii) fails. In particular consider a game in which the utility of the principals and the range of $a$ are as in the previous example, while $b_i \in [0, \frac{1}{2}]$ and $u_0 = \begin{cases} b_1 + b_2 - 2a & \text{if } 0 \leq a \leq \frac{1}{2} \\ b_1 + b_2 - 1 & \text{if } \frac{1}{2} < a \leq 1 \end{cases}$.

Under these assumptions the bidding functions $b_i^o(a) = \begin{cases} a & \text{if } 0 \leq a \leq \frac{1}{2} \\ \frac{1}{2} & \text{if } \frac{1}{2} < a \leq 1 \end{cases}$ are truthful responses to each other relative to $u_i^o = 0$. Additionally, the pair $(a, b^o(\cdot))$ satisfies conditions (Ai) and (Aii) for all $a \in [0,1]$ and condition (Bi) for all $a \in [0, \frac{1}{2}]$. However, condition (Bii) fails because for example, for $a = 1$, $u_i(1, b_i^o(1)) = 0.5 > u_i^o = 0$. Following this failure, the bidding functions $b^o(\cdot)$ do not yield a truthful equilibrium. In order to verify this fact, consider a deviation by one of the principals to the truthful biding function with respect to $u_i = 0.5$.

Proposition 3A and especially proposition 3B are very helpful in the calculation of truthful equilibria. In this respect, I provide examples in section 6 and appendix B.3.

Now I turn to efficiency.

**Proposition 4.** *Efficiency*

Consider a game which exhibits conflict of interests and weak deep pockets at $(a^o, b^o(\cdot))$, which is a truthful equilibrium of this game. Then, there does not exist a feasible pair $(a^*, b^*)$, such that $u(a^*, b^*) \geq u(a^o, b^o(a^o))$ and $u_0(a^*, b^*) \geq u_0(a^o, b^o(a^o))$, with at least one strict inequality.

Proof: See appendix A

Proposition 4 states that under certain conditions, truthful equilibria implement an allocation which is Pareto efficient for all participants of the game (principals and agent). Definition 7 and proposition 5 below summarize the results so far.

**Definition 7.** *Validity*

A truthful equilibrium of a game is valid, if it satisfies propositions 3 and 4 and the game satisfies proposition 2.

**Proposition 5.** *Results*

**a)** All truthful equilibria are valid in all games that satisfy one of the following:

**(i)** lobbying monotonicity and no externalities

**(ii)** market monotonicity and no externalities[8]

---

[8] For similar results in models with quasi linear utility functions see also Bernheim and Whinston (1986), Laussel and Le Breton (2001), Martimort and

**(iii)** lobbying monotonicity and small externalities[9]

**b)** Symmetric truthful equilibria are valid in games that satisfy symmetric negative externalities and lobbying monotonicity.

Proof: Follows directly from propositions 1-4.

Part (a-i) of proposition 5 restates the argument by Dixit et al. (1997) in favour of truthful equilibria in the standard model, while parts (a-ii), (a-iii) and (b) generalize this argument in different settings. The intuition behind proposition 5 is straightforward for the case of market monotonicity and no externalities. This case is the reverse of the model by Dixit et al. (1997) and thus its motivation is similar. In the case of small externalities the intuition is also simple. Specifically, if the externalities are small they have no effect. I discuss this issue further in section 5. For the case of symmetric externalities I provide examples that highlight the role of both symmetry and quasi-concavity in achieving conflict of interests, in appendix B.2.2.

Proposition 5 describes 4 general settings in which truthful equilibria are relevant. However, other such settings might also exist. In this respect, interested researchers can check whether a specific application satisfies any or all of propositions 2-4 directly.

In section 6, I discuss economic economic applications of proposition 5. Now, I turn to a variation of the main model.

## 4 Private common agency

The analysis so far implies that the principals condition their bids on the entire agent's action. For example, in a market game, the principals condition their selling prices, both on the quantity they sell themselves to the agent, but also on the quantity all other principals sell to the agent. In section 6, I consider an example

---

Stole (2003), Segal and Whinston (2003) and Chiesa and Denicolò (2009).

[9] Dasgupta and Maskin (2000) also find that small externalities lead to efficient equilibria in their study of efficient equilibria in Vickrey auctions. I am indebted to professor de Frutos for pointing that to me.

which is consistent with this assumption. However, in many cases the sellers cannot observe the trade between other parties. Moreover, government regulations might forbid the use of such information in contracts. For these reasons, I consider here a variation of the game, named private common agency.

In this modified version of the game, the agent's action $a$ is a vector of dimension $(n)$, which is equal to the number of principals. Formally,

$$a = (a_1, a_2, ..., a_i, ..., a_n) \in A = \times A_i \subset R^n.$$

Moreover, the bidding functions of the principals take the form:

$$b_i : A_i \to R \quad \text{such that} \quad b_i = b_i(a_i)$$

and the utility function of principal $i$ is:

$$u_i : A_i \times R \to R \quad \text{such that} \quad u_i = u_i(a_i, b_i).$$

Finally, the two functions that define feasibility are:

$$\overline{b_i} : A_i \to R \quad \text{and} \quad \underline{b_i} : A_i \to R.$$

The rest of the setting is as in section 2.

I call the game defined above a **private game**. In this new setting the definitions of all other concepts that appear in section 2 must be modified appropriately. These modifications are straightforward, so I omit them and jump directly to the main result.

**Proposition 6.** *Validity in private games*
All truthful equilibria in private games that exhibit opposing monotonicity are valid.
Proof: See appendix B.1.4

Chiesa and Denicolò (2009), also study private games and find that all equilibria are efficient and that truthful equilibria satisfy proposition 3A. Their setting differs from mine in two ways. First, Chiesa and Denicolò (2009) assume that the principals also choose a subset of $A_i$ on which they condition their bids. Second, in their model, these authors allow only for quasi linear utility functions.

I turn now to a further discussion of efficiency.

# 5 Discussion

In this section I discuss games with externalities that do not satisfy propositions 2-4. Specifically, I investigate the existence of efficient equilibria in such games. As it turns out, efficient equilibria might or might not exist depending on the specifics of the model. I discuss this issue with the help of two examples. These examples also clarify the relationship between the paper in hand and the existing literature on common agency with externalities. I start with an example in which there are no efficient equilibria.

**Example 1**

Assume there are two principals ($i = 1, 2$) with utility functions $u_i = a - b_i + \gamma b_j$, in which $a \in [0, 1]$ is the agent's action, $b_i \in [0, a]$ is the bid of principal $i$ and $\gamma$ is a positive parameter. The utility of the agent is $u_0 = b_1 + b_2$.

First, I assume $\gamma = 2$. In this case, the example satisfies opposing monotonicity, but violates small externalities. Moreover, as I show in appendix B.2.1 there exists only one allocation, which is both symmetric and efficient. In this allocation: $a = 1$, $b_1 = b_2 = 1$ and $u_0 = u_1 = u_2 = 2$. Henceforth, I call this allocation, allocation A. However, allocation A is not an equilibrium. To see this, consider any bidding functions such that $b_i(1) = 1$ for both $i$ and think of the following deviation by principal 1: $b_1 = 0,5$ if $a = 1$ and $b_1 = 0$ for all $a \neq 1$. Then, no matter what is the bidding function of principal 2, the agent chooses $a = 1$ and total bids decrease. Moreover, as I also show in appendix B.2.1, there are no efficient equilibria in this example. On the contrary, there is a unique symmetric inefficient equilibrium which is intuitive[10]. Specifically, consider the following allocation, which I name allocation B: $a = 1$, $b_1 = b_2 = 0$. This allocation yields $u_0 = 0$, $u_1 = u_2 = 1$ and can be supported as an equilibrium, by the constant bidding functions $b_i(a) = 0$ for all $a \in [0, 1]$.

Allocations A and B illustrate the cause of efficiency failure in common agency with externalities. This cause is a prisoners' dilemma. Specifically, the princi-pals can benefit from committing to high bids, as in allocation A. However, this commitment is not viable. This is so, because each principal has the motive to unilaterally deviate from any such "agreement" and offer smaller bids. This deviation generates a race to the bottom which leads to the inefficient equilibrium B.

---

[10] Also, there exist many asymmetric inefficient equilibria that are supported by implausible off equilibrium strategies. I provide an example in appendix B.2.1.

This prisoners' dilemma also characterizes the examples provided by Peters (2001) and Martimort and Stole (2002). These authors, observe that externalities in common agency models might lead to inefficient equilibria. Following this observation, they suggest an extension of the common agency model. This extension, allows bids to depend on the biding functions of the other principal and restores efficiency. This idea in terms of example 1 is as follows: Allocation A is not an equilibrium because principal 1 deviates. However, if the bids of one principal depend on the biding function of the other, then principal 2 can use his bidding function to punish the deviating behaviour and support the equilibrium.

In contrast to this approach, I bypass the solution of the efficiency issue altogether. Instead, I notice that in certain cases a prisoners' dilemma does not appear. Indeed, if $\gamma = 0.5$, example 1 satisfies proposition 5. Thus, if a truthful equilibrium exists it is efficient. Specifically, allocation B is such an equilibrium. This result, follows from the fact that if $\gamma = 0.5$, the principals can not benefit from high bids. In general, small externalities do not allow for prisoner dilemmas and therefore lead to efficient equilibria. A similar argument holds for symmetric negative externalities.

Now I turn to an example which violates proposition 5, but nevertheless has an efficient equilibrium.

**Example 2**

Consider a variation of example 1 in which the utility of the agent is $u_0 = -(b_1 + b_2)$ and $\gamma = 2$. This example violates both opposing monotonicity and small externalities. Yet, allocation B can still be supported as an efficient equilibrium, by the constant bidding functions $b_i(a) = 0$ for all $a$.

In this case, allocation B is efficient because the agent dislikes bids. In any other allocation with $a = 1$ and positive bids, the principals might be better off but the position of the agent deteriorates. This is in contrast to example 1, in which both principals and agent get worse off, as a result of a decrease in bids. In example 2, the prisoners' dilemma between principals remains, however it does not hinder efficiency, due to the characteristics of the agent's utility function.

Yet, although allocation B is an efficient equilibrium it is not also a truthful one. This is so, because the structure of truthful bidding functions renders them meaningless without opposing monotonicity. Indeed, in such cases truthful bidding functions fail, since they imply that the principals offer to the agent something he dislikes. Boultzis (2015), considers an example of this situation.

Examples 1 and 2 above discuss the role of small externalities and opposing monotonicity in the efficiency of equilibria. In appendix B.2.2, I provide some additional examples that outline the role of symmetry and quasi-concavity.

Next, I consider economic applications.

# 6 Applications

In this section, I provide two economic applications of proposition 5.

## 6.1 Market application

This is a case of market monotonicity and no externalities.

Assume $n = 2$. The two principals are sellers who sell a homogeneous good. The agent is a buyer who has decided to buy $\overline{q}$ units of the good. For example, think of a government which is in the market for a fixed number of warships.

First, the two sellers submit an offer for a unit price that depends on the quantity that the buyer buys from each seller. Then, the buyer decides how much to buy from each seller. This model is known in the literature as split-award procurement[11]. Although in this model each seller conditions his price exclusively on the quantity he sells to the buyer it is still a public common agency model. This is so because the buyer's choice is essentially one dimensional. In particular when the agent chooses to buy quantity $q_1$ from principal 1, this decision also determines the quantity he buys from principal 2 ($q_2 = \bar{q} - q_1$).

---

[11] See Anton and Yao (1989) and Chiesa and Denicolò (2009).

The profit function of seller $i$ is:

$$\Pi_i = p_i q_i - c q_i^2.$$

Here, $p_i$ is the unit price, $q_i$ stands for quantity and $c > 0$ is a cost parameter.

The buyer wants to allocate $\bar{q}$ between the two sellers in a way that minimizes his expenditure. Thus, the utility of the buyer is:

$$u_0 = -p_1 q_1 - p_2(\bar{q} - q_1).$$

The reservation price for the buyer is $\bar{p}$. You can think of the reservation price as the cost of not buying a unit or as the unit price offered by an outside source.

The bounds that define feasibility for prices are :

$$\underline{p_i}(q_i) = c q_i \geq 0 = p_{min} \quad \text{and} \quad \overline{p_i}(q_i) = \bar{p}.$$

These conditions reflect the fact that prices must be positive, must not be less than the average cost and must not exceed the reservation price. Furthermore, assume $\bar{p} > 3c\bar{q}$.

This model satisfies market monotonicity and lacks externalities. Therefore it satisfies proposition 5. As I show in appendix B.3.1, solving for a truthful equilibrium yields the following symmetric outcome:

$$q_i = \frac{\bar{q}}{2}, \quad p_i = \frac{3c\bar{q}}{2}, \quad \Pi_i = \frac{c\bar{q}^2}{2}$$

while the equilibrium price functions are

$$p_i(q_i) = \begin{cases} \frac{c\bar{q}^2}{2q_i} + c q_i & \text{if} \quad q_i \geq \frac{\bar{p} - \sqrt{\bar{p}^2 - 2c^2\bar{q}^2}}{2c} \\ \bar{p} & \text{if} \quad q_i < \frac{\bar{p} - \sqrt{\bar{p}^2 - 2c^2\bar{q}^2}}{2c} \end{cases}$$

In this model, total cost decreases with the number of producers, because of the increasing marginal cost. Specifically, the total cost of producing quantity $\bar{q}$ of the good is $c\bar{q}^2$ if there is only one producer, while it is cut in half, if there are two producers. Consequently, a familiar result emerges.

Specifically the profits of each principal equal the decrease in the total cost due to his participation in production. This result is standard for truthful equilibria in common agency games. In this respect, Bergemann and Välimäki (2003) show that in such cases, each principal receives his contribution to the social surplus[12].

I turn now to an application with externalities.

## 6.2 Lobbying and public goods

I consider an example with lobbying monotonicity and symmetric negative externalities. In particular, I introduce lobbying to a local public goods model inspired by Persson and Tabellini (1994).

Consider a federal country with $n$ identical states, each populated by one individual. In this country, there is a federal government with the sole purpose to redistribute income across states. This government imposes a tax or subsidy $t_i$ to each of the states. These taxes satisfy:

$$\sum_i t_i = 0.$$

Each state offers to the federal government a bribe $b_i$ in order to affect the choice of $t_i$. In this respect the utility of the federal government is given by:

$$u_0 = \sum_i b_i.$$

Moreover, there are two goods in the economy. These goods are the private good $c$ and the public good $G$. Each state finances the public good by offering a voluntary contribution $g_i$. Thus the total amount of public good equals:

$$G = \sum_i g_i.$$

An example of such a public good is public safety. Each state decides independently how much to spend on the security of its airport. However, since terrorists who arrive in one state can move freely to all states, this spending affects safety in the whole country.

---
[12] On the structure of payoffs in such games see also Laussel and Le Breton (2001) and Villemeur and Versaevel (2003).

Total spending in each state is financed by an endowment $e$. This endowment plus or minus the federal transfer $t_i$ is used for private good consumption, contribution to the public good and bribes. Therefore, the budget constraint for each state is:
$$e + t_i = c_i + b_i + g_i,$$
where $t_i \in [-e, e]$. Finally the utility in state $i$ is :
$$u_i = c_i G.$$

The timing in this model is as follows: First, a common agency game determines bribes and transfers. Then, the two states decide on their public good contributions simultaneously[13].

Following this timing, I start by determining the equilibrium public good contributions. Define, $e_i = e + t_i - b_i$. Then, the unique Nash equilibrium yields $g_i = e_i - \frac{\sum_i e_i}{n+1}$ and $G = \frac{\sum_i e_i}{n+1}$, which implies that:
$$u_i = \left(\frac{\sum_i e_i}{n+1}\right)^2$$

This function is obviously concave and symmetric with respect to $b_i, b_j$. Moreover, $b_i$ is appropriately bounded since:
$$0 \leq b_i \leq e + t_i \leq 2e.$$

This chain of inequalities implies that $\overline{b_i}(\cdot) = \overline{b_j}(\cdot)$ and $\underline{b_i}(\cdot) = \underline{b_j}(\cdot)$.

Moreover, the derivatives of interest are as follows:
$$\frac{\partial u_0}{\partial b_i} = 1 > 0$$
$$\frac{\partial u_i}{\partial b_i} = -2G < 0$$
$$\frac{\partial u_i}{\partial \tilde{b}_{-i}} = -2G < 0 \quad \text{negative externality}$$

---

[13] Bergemann and Välimäki (2003) and Bhaskar and To (2004) also discuss multi stage models with a separate common agency stage.

Thus, this game satisfies lobbying monotonicity and negative externalities. Furthermore it is symmetric and quasi concave and therefore exhibits symmetric negative externalities. Hence, truthful equilibria are valid. Solving for a truthful equilibrium yields:

$$b_1 = b_2 = 0 \quad \text{and} \quad t_i \in [-e, e] \quad s.t. \sum_i t_i = 0.$$

Specifically, the functions $b_i = 0$ for all feasible $t_i$ can be supported as truthful responses to each other, relative to the equilibrium utility level $u_i = \left(\frac{ne}{n+1}\right)^2$. Moreover, these functions trivially satisfy conditions (Ai) and (Aii). Finally, when bribes equal their maximum value $(e + t_i)$ the utility of the principals always equals zero which is its lower bound. Thus, this application satisfies strong deep pockets, which implies that the suggested allocations also satisfy condition (Bii) and therefore constitute truthful equilibria.

In this model the agent has no "bargaining power". Thus, as in similar models without externalities he ends with nothing. Specifically, since the utility of each state depends on total income, the government can not affect it by redistributive transfers. Thus, there is no room for bribes.

## 6.3 Other applications of proposition 5

The two applications above belong in two of the general settings identified by proposition 5. I consider them here because they are simple and characteristic of how proposition 5 can be used. However, proposition 5 considers two more settings, which I briefly review in the remaining of this subsection.

The first setting incorporates applications which exhibit lobbying monotonicity and no externalities. This is the case discussed by Dixit et al. (1997). This type of models is often used to describe situations associated with lobbying. I do not discuss them any further, since they have been studied extensively in the literature[14].

---

[14] Early examples of this literature are Grossman and Helpman (1994), Dixit (1996), Persson (1998), Mitra (1999) e.t.c. More recent examples are Cam-

The second setting includes models with lobbying monotonicity and small externalities. Such situations occur in lobbying models, in which the relation between the principals goes beyond sharing a common agent, very much like in the application in 6.2 above. For example, consider a variation of this application in which the states have different endowments and the good $G$ is not a pure public good. In this respect, assume that $G_i = g_i + \gamma \sum_{i \neq j} g_j$, where $G_i$ is the quantity of good $G$ consumed by state $i$ and $\gamma \in (0, 1)$ is a parameter capturing the externality between states. This modified application exhibits lobbying monotonicity and small externalities[15].

I turn now to the concluding remarks.

# 7 Conclusions

In this paper I provide a set of conditions under which truthful equilibria are valid in common agency models. These conditions generalize the work of Dixit et al. (1997) and Bernheim and Whinston (1986) on this issue. Furthermore, I identify two new families of economic applications to which these conditions apply. In this regard, this paper shows that the scope of truthful equilibria is broader than believed so far.

In terms of future research my results can be useful in two ways. First, the conditions listed in proposition 5 apply in a wide variety of economic models. The validity of truthful equilibria in these models provides for a simple and intuitive way of solving an otherwise difficult problem. Second, propositions 2-4 can help identify additional settings in which truthful equilibria are valid in the future. Thus, the scope of truthful equilibria might be extended even further.

---

pante and Ferreira (2007), Aidt and Hwang (2008) and Esteller-Moré et al. (2012) among many others. For a related work see also Felli and Merlo (2006).

[15] Additional details are available upon request

# A  Appendix

### Proof of proposition 4

Assume the contrary is true. Then, there exists a feasible pair $(a^*, b^*)$ and a truthful equilibrium of the game $(a^o, b^o(\cdot))$, such that $u_0(a^*, b^*) \geq u_0(a^o, b^o(a^o))$ and $u(a^*, b^*) \geq u(a^o, b^o(a^o))$, with at least one strict inequality. These inequalities yield:

$$u_0(a^*, b^*) \geq u_0(a^o, b^o(a^o)) \geq u_0(a^*, b^o(a^*))$$

and

$$u(a^*, b^*) \geq u(a^o, b^o(a^o)) \geq u(a^*, b^o(a^*)).$$

The last inequality in the first expression above holds because $(a^o, b^o(\cdot))$ is an equilibrium. Also, the last inequality in the second expression is due to deep pockets.

The two chains of inequalities above imply that $u_0(a^*, b^*) \geq u_0(a^*, b^o(a^*))$ and $u(a^*, b^*) \geq u(a^*, b^o(a^*))$, with at least one strict inequality. Then, there is a contradiction. Indeed, if $u_0(a^*, b^*) > u_0(a^*, b^o(a^*))$ it follows from conflict of interests that there is an $i$, such that $u_i(a^*, b^*) > u_i(a^*, b^o(a^*))$, which contradicts the second chain of inequalities. Alternatively, if $u(a^*, b^*) > u(a^*, b^o(a^*))$ conflict of interests implies that $u_0(a^*, b^*) > u_0(a^*, b^o(a^*))$, which contradicts the first chain of inequalities. Q.E.D.

# B  Online Appendix

## B.1  Proofs

### B.1.1  Proof of proposition 1

**Proof of (i)**  I show that the first version of strong deep pockets (D.2.1) implies weak deep pockets (D.1). Let $(a^o, b^o(\cdot))$ be a truthful equilibrium and $u^o = u(a^o, b^o(a^o))$. Also, let $(a^*, b^*)$ be a feasible pair. Because of strong deep pockets, $u_i^o \geq u_i(a, \overline{b_i}(a), b_{-i}^o(a))$ for all $i$ and $a \in A$. Then, the definition of truthful bidding functions implies either a) $b_i^o(a^*) = \phi_i(a^*; u^o, b_{-i}^o(\cdot))$ or b) $b_i^o(a^*) = \underline{b_i}(a^*)$. If (a) holds, then $u_i(a^*, b^o(a^*)) = u_i^o$, while if (b) holds, then $u_i(a^*, b^o(a^*)) < u_i^o$. Taking (a) and (b) together yields $u_i^o = u_i(a^o, b^o(a^o)) \geq u_i(a^*, b^o(a^*))$, for all $i$. Since this inequality holds for all feasible pairs, it also holds for the feasible pair in the definition of weak deep pockets. The proof that the second version of strong deep pockets implies weak deep pockets is essentially the same. Q.E.D.

**Proof of (ii)**  I consider market monotonicity. The proof for lobbying monotonicity is essentially the same. The fact that market monotonicity implies conflict of interests in models without externalities is obvious. In what follows I prove that it also implies weak deep pockets.

Let $(a^o, b^o(\cdot))$ be a truthful equilibrium and $u^o = u(a^o, b^o(a^o))$. Then, following the definition of weak deep pockets consider a feasible pair $(a^*, b^*)$ such that $u(a^*, b^*) \geq u(a^o, b^o(a^o)) = u(a^*, \phi(a^*; u^o))$. Because $u(\cdot)$ is increasing in own bids it must be that $b_i^* \geq \phi_i(a^*; u^o)$ for all $i$. Furthermore, since $b_i^*$ is feasible it follows that $b_{max} \geq b_i^* \geq \underline{b_i}(a^*)$. Therefore, $b_{max} \geq \phi_i(a^*; u^o)$. This last inequality and the definition of truthful responses imply that either $b_i^o(a^*) = \phi_i(a^*; u^o)$ or $b_i^o(a^*) = \underline{b_i}(a^*)$. In turn, these equalities along with the feasibility of $b_i^*$ imply that $b_i^* \geq b_i^o(a^*)$ for all $i$. Then, the fact that $u_0(\cdot)$ is decreasing in all bids yields: $u_0(a^*, b^o(a^*)) \geq u_0(a^*, b^*) \geq u_0(a^o, b^o(a^o))$. If any of these two inequalities is strict then we have a contradiction since $(a^o, b^o(\cdot))$ is an equilibrium. Thus, $u_0(a^*, b^o(a^*)) = u_0(a^*, b^*) = u_0(a^o, b^o(a^o))$. Now, if there exists an $i$ such that $u_i(a^*, b_i^o(a^*)) > u_i^o$ we arrive at a contradiction since the pair $(a^o, b^o(\cdot))$ vio-lates trivially the definition of equilibrium. Indeed, in such a case it follows from $u_0(a^*, b^o(a^*)) = u_0(a^o, b^o(a^o))$ that $a^* \in arg\max_{a \in A} u_0(a, b_i^o(a), b_{-i}^o(a))$. Thus, $a^o$ can not be part of an equilibrium. Therefore, $u(a^o, b^o(a^o)) \geq u(a^*, b^o(a^*))$. Q.E.D.

**Proof of (iii)**  **Part 1.**

First, I show that lobbying monotonicity and small externalities imply conflict of interests.

The total differential of the utility of the principals is $du_i = \frac{\partial u_i}{\partial b_i} db_i + \frac{\partial u_i}{\partial \tilde{b}_{-i}} d\tilde{b}_{-i}$. Rearranging terms yields: $du_i = (\frac{\partial u_i}{\partial b_i} - \frac{\partial u_i}{\partial \tilde{b}_{-i}}) db_i + \frac{\partial u_i}{\partial \tilde{b}_{-i}} d\tilde{b}$.

In order to proceed I need to show two things. First, that when the agent's utility increases ($d\tilde{b} > 0$), there is at least one principal who becomes worse off ($du_i < 0$ for at least one $i$).

I start with negative externalities ($\frac{\partial u_i}{\partial \tilde{b}_{-i}} < 0$). In this case small externalities imply that $\frac{\partial u_i}{\partial b_i} - \frac{\partial u_i}{\partial \tilde{b}_{-i}} < 0$. Then, since $d\tilde{b} > 0$ there is at least an $i$ such that $db_i > 0$. As a result, $du_i < 0$ for at least one principal.

Now, I turn to positive externalities ($\frac{\partial u_i}{\partial \tilde{b}_{-i}} > 0$). Consider the principal with the largest increase in bids. Then, for this principal $ndb_i \geq d\tilde{b}$. Substituting this expression in the expression for $du_i$ yields $du_i \leq (\frac{\partial u_i}{\partial b_i} + (n-1)\frac{\partial u_i}{\partial \tilde{b}_{-i}}) db_i < 0$. The last inequality follows from small externalities, since $\frac{\partial u_i}{\partial b_i} + (n-1)\frac{\partial u_i}{\partial \tilde{b}_{-i}} < 0$.

The second thing I need to show is that if all principals are weakly better off and at least one is strictly better off ($du > 0$), then the utility of the agent decreases ($d\tilde{b} < 0$). Assume the contrary, $d\tilde{b} \geq 0$. Then, if $db_i = 0$ for all $i$ it must be that $du_i = 0$ for all $i$, which is a contradiction. Thus, for at least an $i$, it must be that $db_i > 0$. Then, for this principal the analysis above implies that $du_i < 0$, which is a contradiction.

**Part 2.**

In this part, I show that small externalities along with lobbying monotonicity imply weak deep pockets. I start with the proof of a lemma that I will use later during the rest of the proof.

**Lemma 1**

Consider a game that exhibits small negative externalities and lobbying monotonicity. Then, the following is true:

If $b, b'$ are feasible for some $a \in A$ and there exists an $i$, such that $b_i' > b_i$ and $u_i(a, b_i', \tilde{b}'_{-i}) \geq u_i(a, b_i, \tilde{b}_{-i})$ then, $\tilde{b}' < \tilde{b}$.

Proof of lemma 1:

Assume the contrary: $\tilde{b}' \geq \tilde{b} \Rightarrow b_i' - b_i \geq \tilde{b}_{-i} - \tilde{b}'_{-i} > 0$. The last inequality follows from $b_i' > b_i$, $u_i(a, b_i', \tilde{b}'_{-i}) > u_i(a, b_i, \tilde{b}_{-i})$ and negative externalities. Consider $\kappa \geq 0$ and define $u_i(\kappa) = u_i(a, b_i + \kappa, \tilde{b}_{-i} - \kappa)$. Then, $\frac{\partial u_i}{\partial \kappa} = \frac{\partial u_i}{\partial b_i} - \frac{\partial u_i}{\partial \tilde{b}_{-i}} < 0$. The last inequality is due to small negative externalities. Then, using $\frac{\partial u_i}{\partial \kappa} < 0$ yields $u_i(0) > u_i(b_i' - b_i)$ or $u_i(a, b_i, \tilde{b}_{-i}) > u_i(a, b_i', \tilde{b} - b_i') \geq u_i(a, b_i', \tilde{b}'_{-i})$. The last inequality is because by assumption $\tilde{b}' \geq \tilde{b}$ and negative externalities. This last chain of inequalities implies $u_i(a, b_i, \tilde{b}_{-i}) > u_i(a, b_i', \tilde{b}'_{-i})$, which is a contradiction. This concludes the proof of the lemma.

I proceed now with the rest of the proof of part 2.

The properties of $(a^*, b^*)$ and $(a^o, b^o(a^o))$ in the definition of weak deep pockets imply that $u_0(a^*, \tilde{b}^*) \geq u_0(a^o, \tilde{b}^o(a^o)) \geq u_0(a^*, \tilde{b}^o(a^*))$. In turn, this chain of inequalities implies that $\tilde{b}^* \geq \tilde{b}^o(a^*)$. Assume there is an $i$, such that $u_i(a^*, b^o(a^*)) > u_i(a^o, b^o(a^o))$. Then, if $\tilde{b}^* = \tilde{b}^o(a^*)$ it follows that $u_0(a^*, \tilde{b}^*) = u_0(a^o, \tilde{b}^o(a^o)) = u_0(a^*, \tilde{b}^o(a^*))$. In this case there is a contradiction since the pair $(a^o, b^o(\cdot))$ violates trivially the definition of equilibrium. Specifically, it follows from $u_0(a^*, \tilde{b}^o(a^*)) = u_0(a^o, \tilde{b}^o(a^o))$ that $a^* \in \arg\max_{a \in A} u_0(a, \tilde{b}^o(a))$. This last observation, along with $u_i(a^*, b^o(a^*)) > u_i(a^o, b^o(a^o))$ imply that $a^o$ can not be part of an equilibrium. Therefore, $\tilde{b}^* > \tilde{b}^o(a^*)$.

Since the utility of the agent is strictly increasing in total bids, $u_0(a^*, \tilde{b}^*) > u_0(a^*, \tilde{b}^o(a^*))$. Then, following part 1 of this proof, there exists an $i$ such that $u_i(a^*, b^o(a^*)) > u_i(a^*, b^*) \geq u_i(a^o, b^o(a^o))$. For this $i$, the definition of truthful responses implies that $b_i^o(a^*) = \overline{b}_i(a^*) \geq b_i^*$.

I consider two cases. First, I consider positive externalities. Since $u_i(a^*, b^o(a^*)) > u_i(a^*, b^*)$ and $b_i^o(a^*) \geq b_i^*$, it must also be that $\tilde{b}^o_{-i}(a^*) \geq \tilde{b}^*_{-i}$ with at least one strict inequality. However, in this case $\tilde{b}^o(a^*) > \tilde{b}^*$, which is a contradiction.

Second, I turn to negative externalities. In this case, I start by observing that $b_i^o(a^*) \geq b_i^*$ does not hold for all $i$, because if it does, it follows that $\tilde{b}^o(a^*) \geq \tilde{b}^*$, which is a contradiction. Thus, there is at least an $i$, such that $b_i^o(a^*) < b_i^* \leq \overline{b}_i(a^*)$. Then, the definition of truthful responses implies that $u_i(a^*, b^o(a^*)) \leq u_i(a^o, b^o(a^o)) \leq u_i(a^*, b^*)$. Finally, using lemma 1 yields $\tilde{b}^o(a^*) > \tilde{b}^*$, which is a contradiction. Thus $u(a^o, b^o(a^o)) \geq u(a^*, b^o(a^*))$. Q.E.D.

**Proof of (iv)**

**Conflict of interests.**

I prove first that lobbying monotonicity and symmetric negative externalities imply conflict of interests at symmetric allocations.

Consider a game that exhibits lobbying monotonicity and symmetric negative externalities and let $(a, b^*)$, $(a, b)$ be two feasible pairs and also let $(a, b)$ be symmetric. Then I will prove that, (i) if $\tilde{b}^* > \tilde{b}$ there is an $i$ such that $u_i(a, b^*) < u_i(a, b)$ and (ii) if $u(a, b^*) > u(a, b)$ then $\tilde{b}^* < \tilde{b}$.

I start with the proof of (i). I assume the contrary: $u_i(a, b^*) \geq u_i(a, b)$, for all $i$.

Then, because $\tilde{b}^* > \tilde{b}$, there exists $\xi_i \in [-b_i, b_{max} - b_i]$ for all principals $i$, such that $b_i^* = b_i + \xi_i$ and $\sum_i \xi_i > 0$. Moreover, because of the symmetry of the game and $(a, b)$, it follows from $u_i(a, b^*) \geq u_i(a, b)$ that $u_k(a, b_i^*, \tilde{b}_{-i}^*) \geq u_k(a, b_k, \tilde{b}_{-k})$ for all $i$ and $k$. Then, the quasi-concavity of $u_k(\cdot)$ yields $u_k(a, \frac{1}{n}\sum_i b_i^*, \frac{1}{n}\sum_i \tilde{b}_{-i}^*) \geq u_k(a, b_k, \tilde{b}_{-k})$. However, $\frac{1}{n}\sum_i b_i^* = \frac{1}{n}\sum_i b_i + \frac{1}{n}\sum_i \xi_i = b_k + \frac{1}{n}\sum_i \xi_i > b_k$. Also, $\frac{1}{n}\sum_i \tilde{b}_{-i}^* = \frac{n-1}{n}\sum_i b_i + \frac{n-1}{n}\sum_i \xi_i = \tilde{b}_{-k} + \frac{n-1}{n}\sum_i \xi_i > \tilde{b}_{-k}$. Then, because of negative externalities $u_k(a, \frac{1}{n}\sum_i b_i^*, \frac{1}{n}\sum_i \tilde{b}_{-i}^*) < u_k(a, b_k, \tilde{b}_{-k})$, which is a contradiction.

I turn now to (ii). Specifically, I must show that if $u(a, b^*) > u(a, b)$ then $\tilde{b}^* < \tilde{b}$. Assume the contrary, $\tilde{b}^* \geq \tilde{b}$. If $\tilde{b}^* > \tilde{b}$ then the analysis above yields a contradiction. Therefore, $\tilde{b}^* = \tilde{b}$. Using symmetry, $u(a, b^*) > u(a, b)$ implies that $u_k(a, b_i^*, \tilde{b}_{-i}^*) \geq u_k(a, b_k, \tilde{b}_{-k})$ for all $i$ and $k$, with at least one strict inequality. Then, the quasi-concavity of $u_k(\cdot)$ yields $u_k(a, \frac{1}{n}\sum_i b_i^*, \frac{1}{n}\sum_i \tilde{b}_{-i}^*) > u_k(a, b_k, \tilde{b}_{-k})$. However, $\frac{1}{n}\sum_i b_i^* = \frac{\tilde{b}^*}{n} = \frac{\tilde{b}}{n} = b_k$. Also, $\frac{1}{n}\sum_i b_{-i}^* = \frac{(n-1)\tilde{b}^*}{n} = \frac{(n-1)\tilde{b}}{n} = (n-1)b_k = \tilde{b}_{-k}$. These equalities yield $u_k(a, \frac{1}{n}\sum_i b_i^*, \frac{1}{n}\sum_i \tilde{b}_{-i}^*) = u_k(a, b_k, \tilde{b}_{-k})$, which is a con-tradiction.

**Weak deep pockets**

I continue to prove that symmetric negative externalities and lobbying monotonicity imply weak deep pockets. In this regard consider such a game and let $(a^o, b^o(\cdot))$ be a symmetric truthful equilibrium and $(a^*, b^*)$ be a feasible pair such that $u_0(a^*, b^*) \geq u_0(a^o, b^o(a^o))$ and $u(a^*, b^*) \geq u(a^o, b^o(a^o))$ with at least one strict inequality. Then, I will show that $u(a^o, b^o(a^o)) \geq u(a^*, b^o(a^*))$.

I start by assuming the opposite. That is, there exists an $i$ for which $u_i(a^*, b^o(a^*)) > u_i(a^o, b^o(a^o))$.

Consider first the case $u_0(a^*, b^*) > u_0(a^o, b^o(a^o))$. Then, $u_0(a^*, b^*) > u_0(a^o, b^o(a^o)) \geq u_0(a^*, b^o(a^*))$ implies that $u_0(a^*, b^*) > u_0(a^*, b^o(a^*))$. Moreover, for the principal $i$ above, the definition of truthful responses yields $b_i^o(a^*) = \overline{b}_i(a^*)$. Then, symmetry implies that $b_i^o(a^*) = \overline{b}_i(a^*)$ for all $i$. Therefore, $\tilde{b}^o(a^*) \geq \tilde{b}^*$ and $u_0(a^*, b^o(a^*)) \geq u_0(a^*, b^*)$ which is a contradiction. The last inequality follows from the fact that the game is cumulative.

Now I turn to the case $u_0(a^*, b^*) = u_0(a^o, b^o(a^o))$. In this case, if $u_0(a^o, b^o(a^o)) > u_0(a^*, b^o(a^*))$ the analysis above can be repeated. Thus, $u_0(a^*, b^*) = u_0(a^o, b^o(a^o)) = u_0(a^*, b^o(a^*))$. This equality implies that $a^*$ also maximizes the utility of the agent when the principals submit $b^o(\cdot)$. Moreover, because of assuming the contrary in the beginning of the proof, there exists an $i$ such that $u_i(a^*, b^o(a^*)) > u_i(a^o, b^o(a^o))$. The existence of this principal leads to a contradiction, since $a^o$ trivially violates the definition of equilibrium. Q.E.D.

### B.1.2 Proof of proposition 2

**Proof of proposition 2**

I prove proposition 2 under the assumption of market monotonicity. The proof for lobbying monotonicity is very similar.

I will show that if any best response of a principal yields a certain utility level then the truthful response relative to this utility level is also a best response. In this respect, let $b_i^o(\cdot)$ be a best response of principal $i$ to the bidding functions $b^o_{-i}(\cdot)$ of the other principals. Then, there exists $a^o \in arg\max_{a \in A} u_0(a, b^o(a))$, such that there does not exist a feasible pair $(a^*, b^*_i(\cdot))$, such that $u_i(a^*, b^*_i(a^*), b^o_{-i}(a^*)) > u_i(a^o, b^o(a^o))$ and $a^* \in arg\max_{a \in A} u_0(a, b^*_i(a), b^o_{-i}(a))$.

Define, $u^o_i = u_i(a^o, b^o(a^o))$ and the truthful response of principal $i$ to the bidding functions of the other principals relative to $u^o_i$ as $b^T_i(a; b^o_{-i}(\cdot), u^o_i)$. For the shake of simplicity in the remaining of the proof I suppress the other arguments and write $b^T_i(a)$. Finally, I define the set $A' = arg\max_{a \in A} u_0(a, b^T_i(a), b^o_{-i}(a))$.

If $a^o \in A'$ then $b^T_i(a)$ is trivially a best response since $b^T_i(a^o) = b^o_i(a^o)$.

I turn now to the case in which $a^o \notin A'$. Consider any $a' \in A'$. Then, I claim that $u_0(a', b^T_i(a'), b^o_{-i}(a')) > u_0(a', b^o(a'))$. If not, then $u_0(a', b^o(a')) \geq u_0(a', b^T_i(a'), b^o_{-i}(a')) > u_0(a^o, b^T_i(a^o), b^o_{-i}(a^o)) = u_0(a^o, b^o(a^o))$. The strict inequality follows from the fact that $a^o \notin A'$ while $a' \in A'$ and the equality from the definition of $b^T_i(\cdot)$. Thus, $u_0(a', b^o(a')) > u_0(a^o, b^o(a^o))$ which is a contradiction because $a^o \in arg\max_{a \in A} u_0(a, b^o(a))$.

Furthermore, because of market monotonicity the utility of the agent is decreasing in all bids. Therefore, $u_0(a', b^T_i(a'), b^o_{-i}(a')) > u_0(a', b^o(a'))$ implies that $b^T_i(a') < b^o_i(a') \leq \overline{b}_i(a')$. Then, following the definition of truthful responses either $b^T_i(a') = \phi_i(a'; u^o_i, b_{-i}(\cdot))$ or $b^T_i(a') = \underline{b_i}(a')$. The last result combined with the increasing utility of the principals in own bids and the definition of truthful responses yields $u_i(a', b^T_i(a'), b^o_{-i}(a')) \geq u^o_i$ which proves that $b^T_i(\cdot)$ is a best response. Q.E.D.

### B.1.3 Proof of proposition 3

I prove proposition 3 under the assumption of market monotonicity. The proof for lobbying monotonicity is very similar.

**Proof of 3A**

I start with the following lemma.

**Lemma**

A feasible pair $(a^o, b^o(\cdot))$ is an equilibrium if and only if:

(i) $a^o \in arg\max_{a \in A} u_0(a, b^o(a))$

(ii)For all $i$, $(a^o, b_i^o(a^o)) \in arg \max_{(a,b_i)} u_i(a, b_i, b_{-i}^o(a))$ subject to $a \in A$, $b_i = b_i(a)$ for some feasible bidding function $b_i(\cdot)$ and $u_0(a, b_i, b_{-i}^o(a)) \geq \max_{a' \in A} u_0(a', b_{max}, b_{-i}^o(a'))$.

Proof of the Lemma:

Necessity:

Assume that $(a^o, b^o(\cdot))$ is an equilibrium but it does not solve the maximization problem in condition (ii) of the lemma. Then, there exists an $i$ and a feasible pair $(a^*, b_i^*)$ which satisfies the constraints in condition (ii) and yields $u_i(a^*, b_i^*, b_{-i}^o(a^*)) > u_i(a^o, b^o(a^o))$.

In this case though, I can show that $(a^o, b^o(\cdot))$ is not an equilibrium which is a contradiction. In order to show this contradiction I need to prove that there exists a feasible bidding function $b_i^*(\cdot)$ such that $b_i^*(a^*) = b_i^*$ and $a^* \in arg \max_{a \in A} u_0(a, b_i^*(a), b_{-i}^o(a))$.

Since $(a^*, b_i^*)$ satisfies the constraints in condition (ii), there exists a feasible bidding function $\hat{b}_i(\cdot)$ such that $\hat{b}_i(a^*) = b_i^*$. Define the function $\varphi_i : A \to R$ implicitly, through $u_0(a, \varphi_i, b_{-i}^o(a)) = u_0(a^*, b_i^*, b_{-i}^o(a^*))$. The function $\varphi_i(\cdot)$ always exists because the utility of the agent is strictly decreasing in all bids. Furthermore, $\varphi_i(a^*) = b_i^*$.

Moreover, because of condition (ii), $u_0(a^*, b_i^*, b_{-i}^o(a^*)) \geq \max_{a' \in A} u_0(a', b_{max}, b_{-i}^o(a'))$ and therefore $u_0(a, \varphi_i(a), b_{-i}^o(a)) = u_0(a^*, b_i^*, b_{-i}^o(a^*)) \geq \max_{a' \in A} u_0(a', b_{max}, b_{-i}^o(a')) \geq u_0(a, b_{max}, b_{-i}^o(a))$ for all $a \in A$.

Since the agent's utility is strictly decreasing in all bids, this chain of inequalities implies that $\varphi_i(a) \leq b_{max}$, for all $a \in A$. Define $b_i^*(a) = max[\varphi_i(a), \hat{b}_i(a)]$. Due to $\varphi_i(a) \leq b_{max}$ and the fact that $\hat{b}_i(\cdot)$ is feasible, $b_i^*(a) \in [\underline{b_i}(a), b_{max}]$ for all $a \in A$ and therefore $b_i^*(\cdot)$ is also feasible. Furthermore, $\hat{b}_i(a^*) = \varphi_i(a^*) = b_i^*$ and therefore $b_i^*(a^*) = b_i^*$.

Finally, I observe that

$u_0(a^*, b_i^*(a^*), b_{-i}^o(a^*)) = u_0(a^*, b_i^*, b_{-i}^o(a^*)) = u_0(a, \varphi_i(a), b_{-i}^o(a)) \geq u_0(a, b_i^*(a), b_{-i}^o(a))$

for all $a \in A$. The last inequality follows from the definition of $b_i^*(\cdot)$ which implies that $b_i^*(a) \geq \varphi_i(a)$. This chain confirms that $a^* \in arg \max_{a \in A} u_0(a, b_i^*(a), b_{-i}^o(a))$ which concludes the proof for necessity.

Sufficiency:

Assume that $(a^o, b_i^o(a^o))$ is a feasible pair that solves the maximization problem in condition (ii) of the lemma but is not part of an equilibrium and that $b_i^o(\cdot)$ is a feasible bidding function. Then, there exists a feasible pair $(a^*, b_i^*(\cdot))$ such that:

a) $a^* \in arg\max_{a \in A} u_0(a, b_i^*(a), b_{-i}^o(a))$ and

b) $u_i(a^*, b_i^*(a^*), b_{-i}^o(a^*)) > u_i(a^o, b^o(a^o))$

However, in this case $u_0(a^*, b_i^*(a^*), b_{-i}^o(a^*)) \geq u_0(a, b_i^*(a), b_{-i}^o(a)) \geq u_0(a, b_{max}, b_{-i}^o(a))$. The first inequality is due to (a) above, while the second inequality is because $b_i^*(\cdot)$ is feasible and therefore $b_i^*(a) \leq b_{max}$ for all $a \in A$. As a result, $(a^*, b_i^*(a^*))$ satisfies the constraints in condition (ii) of the lemma, which implies that $(a^o, b_i^o(a^o))$ does not solve the maximization problem. This contradiction concludes the proof for necessity and the lemma.

Now I proceed with the rest of the proof.

Let $(a^o, b^o(\cdot))$ be an equilibrium of the game. Consider principal $i$. I will show that the inequality in condition (ii) of the lemma, holds as an equality. Assume the contrary. Then, if the equilibrium bid equals the maximum bid or $b_i^o(a^o) = b_{max}$, the contradiction is obvious. If on the other hand $b_i^o(a^o) < b_{max}$, the pair $(a^o, b_i^o(a^o))$ does not solve the maximization problem in the lemma. Indeed, in this case, there exists $b_i^*$ such that $b_{max} > b_i^* > b_i^o(a^o)$. Then, the feasible pair $(a^o, b_i^*)$ satisfies the inequality constraint in condition (ii) of the lemma because $b_{max} > b_i^*$ and yields $u_i(a^o, b_i^*, b_{-i}^o(a^o)) > u_i(a^o, b^o(a^o))$ because $b_i^* > b_i^o(a^o)$. In order to conclude the proof, I need to show that there exists a feasible biding function $b_i^*(\cdot)$ such that $b_i^*(a^o) = b_i^*$. In this respect consider the function $b_i^* = \begin{cases} b_i^o(a) & \text{if } a \neq a^o \\ b_i^* & \text{if } a = a^o \end{cases}$.

**Proof of 3B** The last argument proves proposition 3A. I turn now to proposition 3B.

I start with a useful lemma.

**Lemma: Maxima inequality.**

Consider the feasible bidding functions $b^o(\cdot)$. Let $a' \in arg\max_{a \in A} u_0(a, b_{max}, b^o(a))$

and $a^o \in arg \max_{a \in A} u_0(a, b^o(a))$. Then, $u_0(a^o, b^o(a^o)) \geq u_0(a', b_{max}, b^o_{-i}(a'))$.

**Proof:** Assume the contrary and consider any $a' \in arg \max_{a \in A} u_0(a, b_{max}, b^o(a))$. Then, $u_0(a', b_{max}, b^o(a')) > u_0(a^o, b^o(a^o)) \geq u_0(a', b^o_i(a'), b^o_{-i}(a'))$. Therefore, because the utility of the agent is strictly decreasing in bids it must be that $b^o_i(a') > b_{max}$, which is not possible because $b^o(\cdot)$ is feasible. Thus, $u_0(a^o, b^o(a^o)) \geq \max_{a \in A} u_0(a, b_{max}, b^o(a))$.

I continue now with the rest of the proof. Assume that there exists a pair $(a^o, b^o(\cdot))$ as in the statement of proposition 3B, which is not a truthful equilibrium. Then, because $(a^o, b^o(\cdot))$ is not an equilibrium, there exists an $i$ and a feasible pair $(a^*, b^*_i(\cdot))$ such that $b^*_i(a^*) = b^*_i$, $u_i(a^*, b^*_i, b^o_{-i}(a^*)) > u^o_i$ and $a^* \in arg \max_{a \in A} u_0(a, b^*_i(a), b^o_{-i}(a))$. Then, the maxima inequality lemma above implies that $u_0(a^*, b^*_i, b^o_{-i}(a^*)) \geq \max_{a \in A} u_0(a, b_{max}, b^o_{-i}(a))$. Define $u^*_i = u_i(a^*, b^*_i, b^o_{-i}(a^*))$ and $b^T_i(\cdot)$, which is the truthful response of $i$ to $b^o_{-i}(\cdot)$ relative to $u^*_i$. Then, $b^T_i(a^*) = b^*_i$. Therefore, $u_0(a^*, b^T_i(a^*), b^o_{-i}(a^*)) \geq \max_{a \in A} u_0(a, b_{max}, b^o_{-i}(a)) = u_0(a^o, b^o(a^o)) \geq u_0(a^*, b^o(a^*))$. For future reference I name this chain "principal chain". The first inequality in this chain is due to the maxima inequality lemma, the equality is because $(a^o, b^o(\cdot))$ satisfies condition (Aii) and the last inequality because it satisfies condition (Ai). The principal chain implies that $u_0(a^*, b^T_i(a^*), b^o_{-i}(a^*)) \geq u_0(a^*, b^o(a^*))$ and because the utility of the agent is decreasing in all bids it follows that $b^*_i = b^T_i(a^*) \leq b^o_i(a^*)$. The last inequality along with the fact that the utility of the principals is increasing in own bids implies that $u_i(a^*, b^o_i(a^*), b^o_{-i}(a^*)) \geq u_i(a^*, b^T_i(a^*), b^o_{-i}(a^*)) = u^*_i > u^o_i$. Thus, $u_i(a^*, b^o_i(a^*), b^o_{-i}(a^*)) > u^o_i$, which yields following the definition of truthful responses $b^o_i(a^*) = \underline{b}_i(a^*)$. Then, because $b^T_i(\cdot)$ is feasible it follows that $b^T_i(a^*) \geq \underline{b}_i(a^*) = b^o_i(a^*)$. Taken together $b^T_i(a^*) \geq b^o_i(a^*)$ and $b^T_i(a^*) \leq b^o_i(a^*)$ imply that $b^T_i(a^*) = b^o_i(a^*)$. As a result, $u_0(a^*, b^T_i(a^*), b^o_{-i}(a^*)) = u_0(a^*, b^o(a^*))$ and therefore all inequalities in the principal chain hold as equalities. Thus, $u_0(a^*, b^T_i(a^*), b^o_{-i}(a^*)) = u_0(a^*, b^o_i(a^*), b^o_{-i}(a^*)) = \max_{a \in A} u_0(a, b_{max}, b^o_{-i}(a)) = u_0(a^o, b^o(a^o)) \geq u_0(a, b^o(a))$ for all $a \in A$. In this chain the last inequality holds because of condition (Ai). It follows that both $a^*$ and $a^o$ maximize $u_0(a, b^o(a))$ and satisfy conditions (Ai) and (Aii). Therefore, $u^o_i = u_i(a^o, b^o(a^o)) \geq u_i(a^*, b^o_i(a^*), b^o_{-i}(a^*)) =$ $u_i(a^*, b_i^T(a^*), b_{-i}^o(a^*)) = u_i(a^*, b_i^*, b_{-i}^o(a^*)) = u_i^*$, which is a contradiction because $u_i^* > u_i^o$. In the last chain the first inequality is because of condition (Bii) and the equalities because $b_i^T(a^*) = b_i^o(a^*)$. Q.E.D.

### B.1.4 Proof of proposition 6

The proof of plausibility and efficiency is the same as in section 2, up to the necessary transformations of bidding and utility functions[16]. However, the proof of calculation requires adjustment. Specifically, the related proof in section 2 follows closely Dixit et al. (1997) and uses a lemma whose proof requires bidding functions to depend on the entire set $A$. For this reason, I provide here a new proof for private games. I only prove 3A, since the proof for 3B is very similar to the proof in section 2.

I consider again the case of market monotonicity. Assume that the pair $(a^o, b^o(\cdot))$ is an equilibrium and that $u^o$ is the vector of equilibrium utilities of the principals. First, I observe that from the maxima inequality lemma above (appropriately adapted) it follows that $u_0(a^o, b^o(a^o)) \geq \max_{a \in A} u_0(a, b_{max}, b_{-i}^o(a_{-i}))$.

I will proceed now to show that this expression holds as an equality. Assume the contrary: $u_0(a^o, b^o(a^o)) > \max_{a \in A} u_0(a, b_{max}, b_{-i}^o(a_{-i}))$. Then, because the utility functions of the principals and agent are continuous with respect to bids, there exists $b_i^* \in R$ such that $b_{max} > b_i^* > b_i^o(a_i^o)$, $u_0(a^o, b_i^*, b_{-i}^o(a_{-i}^o)) > \max_{a \in A} u_0(a, b_{max}, b_{-i}^o(a_{-i}))$ and $u_i(a_i^o, b_i^*) > u_i(a_i^o, b_i^o(a_i^o))$. However, in this case, the pair $(a^o, b^o(\cdot))$ is not an equilibrium. To see this consider the following deviation by principal $i$:

$$b_i = \begin{cases} b_{max} & \text{if } a_i \neq a_i^o \\ b_i^* & \text{if } a_i = a_i^o \end{cases}. \text{ Q.E.D.}$$

---

[16] Except from the adjustment of utility and bidding functions, in the proof of proposition 2 the expression "if $a^o \in A'$" must be substituted by "if there exists an $a \in A'$ such that $a_i = a_i^o$".

## B.2 Appendix to discussion

### B.2.1 Example 1

First I show that the unique symmetric efficient allocation is $a = 1, b_1 = b_2 = 1$. Obviously in any efficient allocation $a = 1$. Assume there exists an efficient allocation such that $b_i < 1$ for both $i$. Then, define $\mu = 1 - max\{b_1, b_2\} > 0$. Consequently, if both the principals increase their bids at $a = 1$ by $\mu$ the utility of both the principals and the agent increases. Thus, in all efficient allocations $a = 1$ and at least for one principal $b_i = 1$. In turn, this point implies that there is only one symmetric efficient allocation in which $a = 1$ and for both principals $b_i = 1$.

Now I show that there are no efficient equilibria. Consider an efficient allocation. In any such allocation $a = 1$ and at least for one principal $b_i = 1$. Without loss of generality assume that $b_1 = 1$. Then there are two cases: $b_2(1) > 0$ and $b_2(1) = 0$. If $b_2(1) = \lambda > 0$ then principal 1 can deviate for example to
$$b_1 = \begin{cases} 1 - \frac{\lambda}{2} & \text{if } a = 1 \\ 0 & \text{if } a \neq 1 \end{cases}$$
. In this case the agent still chooses $a = 1$ and the utility of principal 1 increases regardless of his initial off equilibrium strategy.

If instead $b_2(1) = 0$ principal 1 earns $u_1 = 0$. Then, principal 1 can deviate to $b_1 = 0$ for all $a \in [0, 1]$. In this case, if there exists an $a \in [0, 1]$ such that $b_2(a) > 0$ the utility of principal 1 is $u_1 = a + 2b_2 > 0$. If on the other hand $b_2 = 0$ for all $a \in [0, 1]$ the agent is indifferent among all $a$. Then, any $a > 0$ maximizes the utility of the agent and leaves the principal better off. For example, if the agent chooses $a = 1$, principal 1 earns utility $u_1 = 1 > 0$.

Finally, as an example of an asymmetric inefficient equilibrium consider the following: $b_1 = \begin{cases} 0.2 & \text{if } a = 1 \\ 0 & \text{if } a \neq 1 \end{cases}$, $b_2 = \begin{cases} 0.2 & \text{if } a = 0.2 \\ 0 & \text{if } a \neq 0.2 \end{cases}$. These bidding functions yield $a = 1, b_1 = 0.2, b_2 = 0, u_0 = 0.2, u_1 = 0.8, u_2 = 1.4$.

### B.2.2 Symmetric negative externalities

Examples 1 and 2 highlight the role of symmetry.
**Example 1**

Consider a game with two principals in which $u_0 = b_1 + b_2$, $u_i = a - b_i - 2b_j$, $a \in [0,1]$ and $b_i(0) \in [0,5]$ for both $i$. This game is symmetric and quasi-concave. Define allocation $A$ as $a = 0$ and $b_1 = b_2 = 1$. For this allocation $u_0 = 2$ and $u_1 = u_2 = -3$.

Graph 1 describes this situation. In this graph $I_0, I_1$ and $I_2$ are the indifference curves for the agent and the two principals that go through point $A$. The utility of the principals increases at allocations that lie to the south west of point $A$, while the utility of the agent at allocations that lie to the north east of the same point. In this regard the shaded area $f$ depicts the allocations that make both principal 1 and the agent better off than in $A$. In a similar manner, the shaded area $g$ depicts the same allocations for the agent and principal 2. The fact that the areas $f$ and $g$ intersect only at point $A$ implies conflict of interests.

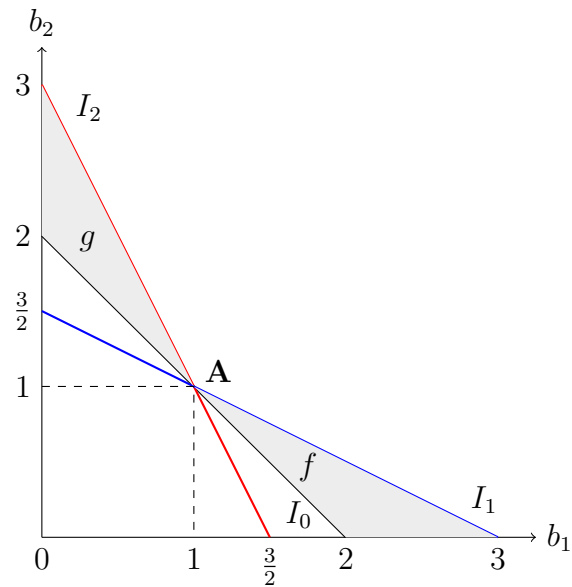

**Graph 1**

Symmetry and quasi-concavity.

**Example 2**

Let me now consider an example in which there is no conflict of interests. Consider a variation of the previous game in which $u_0 = b_1 + b_2$, $u_1 = a - b_1 - 2b_2$ and $u_2 = a - b_1 - 4b_2$. This game is quasi-concave but not symmetric. Define allocation $A$ as $a = 0$, $b_1 = 1$ and $b_2 = 1$. This allocation yields $u_0 = 2$, $u_1 = -3$ and $u_2 = -5$.

Graph 2 reproduces graph 1 for this example. However, in this case, because of asymmetry, there are allocations that make both principals and the agent better off than in $A$. The shaded area in graph 2 depicts these allocations. The existence of such allocations violates conflict of interests.

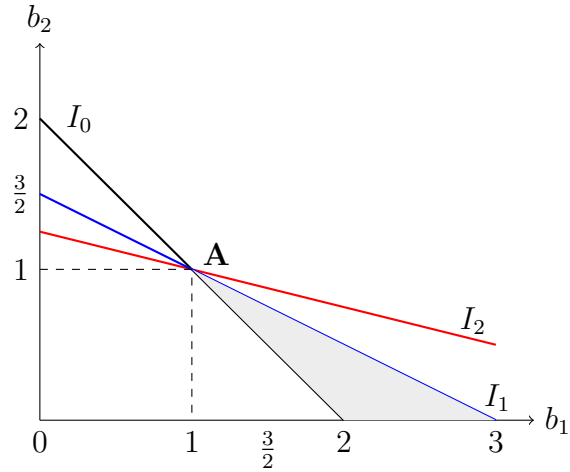

**Graph 2**

Quasi-concavity without symmetry.

Now I turn to quasi-concavity. Examples 3 and 4 that follow highlight its role.

**Example 3**

Consider a game in which $u_0 = b_1 + b_2$, $u_1 = a - b_1 - b_2^2$, $u_2 = a - b_2 - b_1^2$, $a \in [0, 1]$ and $b_i(0) \in [0, 3]$ for both $i$. This game is both symmetric and quasi-concave. Define allocation $A$ as $a = 0$ and $b_1 = b_2 = 1$, which yields $u_0 = 2$, $u_1 = -2$ and $u_2 = -2$. Then, graph 3 depicts the respective indifference curves. As in graph 1 the shaded areas $f$ and $g$ depict the allocations that make the agent and one of the principals better off than in $A$. Again, the fact that $f$ and $g$ intersect only in $A$, implies conflict of interests.

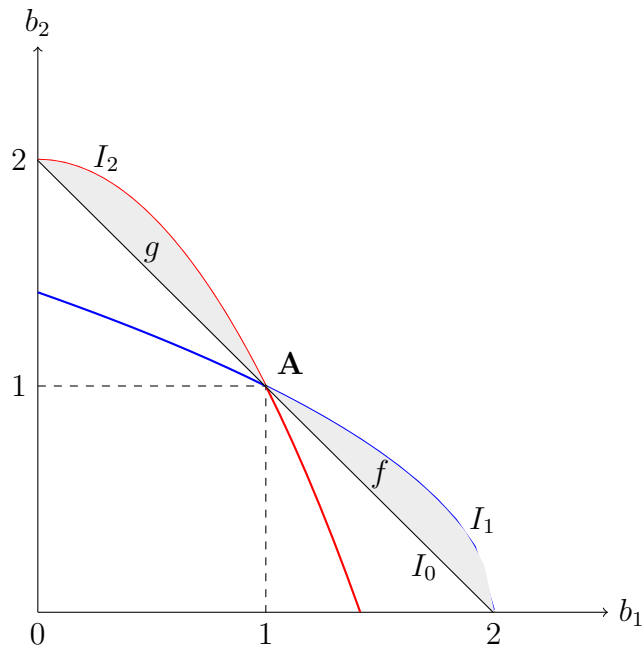

**Graph 3**

Symmetry and quasi-concavity (2).

I turn now to example 4.

**Example 4**

Consider a variation of example 3 in which $u_0 = b_1 + b_2$, $u_1 = a - b_1 - 2\sqrt{b_2}$ and $u_2 = a - b_2 - 2\sqrt{b_1}$. This game is symmetric but not quasi concave. Graph 4 reproduces graph 3 for this example. The shaded area depicts the set of allocations that make both the principals and the agent better off than in $A$. The fact that this set is not empty violates conflict of interests.

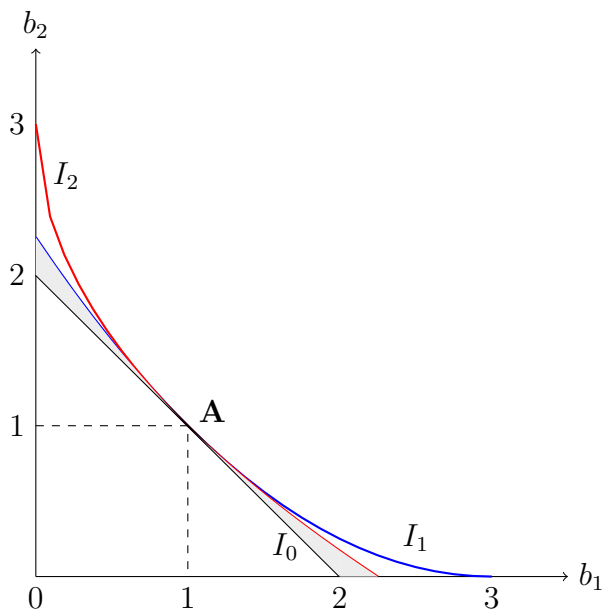

**Graph 4**

Symmetry without quasi-concavity.

## B.3 Appendix to applications

### B.3.1 Market application

In order to solve this problem I use proposition 3B. First I observe that the lower bound of profits is zero. This bound is always reached when the price also takes its lower bound which is equal to the average cost. Thus, this game satisfies strong deep pockets and consequently condition (Bii). Therefore, an allocation of profits ($\Pi_i$) and quantities ($q_i$) which satisfies conditions (Ai) and (Aii) gives rise to a truthful equilibrium.

I proceed by solving the profit function with respect to $p_i$ which yields: $p_i = \frac{\Pi_i + c q_i^2}{q_i} \equiv \varphi(q_i; \Pi_i)$. I guess that this function is the equilibrium price function around the equilibrium quantities. In such a case, when both sellers offer this price function the agent faces the following problem:

$$\max_{q_1 \in [0, \bar{q}]} -p_1 q_1 - p_2(\bar{q} - q_1)$$

The unique solution to this problem is $q_i = \frac{\bar{q}}{2}$, which is independent of $\Pi_i$. Thus for this quantity condition (Ai) is satisfied.

I turn now to condition (Aii) which I use to calculate the equilibrium $\Pi_i$. First, I notice that because of symmetry $\Pi_i = \Pi_j \equiv \Pi$. Then, I guess that when seller $j$ submits the reservation price the buyer buys all the quantity from principal $i$. Under this guess, condition (Aii) becomes $-2\Pi - c\frac{\bar{q}^2}{2} = -\Pi - c\bar{q}^2$, which in turn yields $\Pi = \Pi_i = c\frac{\bar{q}^2}{2}$. I substitute $\Pi_i$ in $\varphi_i(\cdot)$ and get $\varphi(q_i) = \frac{c\bar{q}^2}{2q_i} + cq_i$. Then, in order to verify my first guess I need to show that $c\frac{\bar{q}}{2} \leq \varphi(\frac{\bar{q}}{2}) \leq \bar{p}$ and in order to verify the second that $\varphi(\bar{q}) \leq \bar{p}$.

I proceed, by solving the equation $\bar{p} = \frac{c\frac{\bar{q}^2}{2} + c(q_i^*)^2}{q_i^*}$ with respect to $q_i^*$ in order to calculate the quantity $q_i^*$ at which the price hits the upper bound. This exercise yields: $q_i^* = \frac{\bar{p} - \sqrt{\bar{p}^2 - 2c^2\bar{q}^2}}{2c}$. In this way I obtain the price function $p_i(\cdot)$ that I provide in the main text. Moreover, I find that $\varphi(q_i) < \bar{p}$ for all $q_i > q_i^*$. Then, the assumption $\bar{p} > 3c\bar{q}$ implies that $\frac{\bar{q}}{2} > q_i^*$ and $\varphi_i(q_i) \geq cq_i$ for all $q_i \in [0, \bar{q}]$. These results verify both my guesses and confirm that the proposed equilibrium price function satisfies proposition 3B.